\documentclass[aip, jap, amsmath,amssymb, reprint]{revtex4-1}
\usepackage{amsmath}

\usepackage{graphicx}
\usepackage{dcolumn}
\usepackage{bm}

\begin{document}
\preprint{AIP/123-QED}

\title{Coplanar waveguide based ferromagnetic resonance in ultrathin film magnetic nanostructures: impact of conducting layers}

\author{H. G\l owi\'nski}
\affiliation{Institute of Molecular Physics, Polish Academy of
Sciences, PL-60179 Pozna\'{n}, Poland}
\author{M. Schmidt}
\affiliation{Institute of Molecular Physics, Polish Academy of
Sciences, PL-60179 Pozna\'{n}, Poland}
\author{I. Go\'{s}cia\'{n}ska}
\affiliation{Faculty of Physics, A.Mickiewicz University, Umultowska 85, PL-61614 Pozna\'{n}, Poland}
\author{J. Dubowik}
\affiliation{Institute of Molecular Physics, Polish Academy of Sciences, PL-60179 Pozna\'{n}, Poland}
\email{dubowik@ifmpan.poznan.pl}\thanks{Corresponding author}
\author{J-Ph. Ansermet}
\affiliation{Institute of Condensed Matter Physics, Station 3, Ecole Polytechnique F\'{e}d\'{e}rale de Lausanne — EPFL,
CH-1015 Lausanne, Switzerland}

\date{\today}

\begin{abstract}
We  report broadband ferromagnetic resonance (FMR) measurements based on a coplanar waveguide (CPW) of ultrathin magnetic film structures that comprise in-plane/out-of-plane decoupled layers deposited on nonmagnetic buffer layers of various thickness or other buffer structures with a diverse sheet resistance. We show that the excitation of the fundamental mode can be substantially (up to 10 times) enhanced in the structures deposited on  buffer layers with a low sheet resistance in comparison to the structures deposited on thin or weakly conducting buffer layers. The results are analyzed in terms of shielding of the electromagnetic field of CPW by the conducting buffer layers. The effect of  enhancement of FMR absorption can be attractive for applications in spintronic devices that utilize magnetization dynamics of ultrathin ferromagnetic layers.

\end{abstract}

\pacs{76.50.+g, 75.40.Gb, 75.40.Mg}

\maketitle

\section{Introduction}\label{s1}
Ferromagnetic resonance (FMR) based on a vector network analyzer (VNA) and a coplanar waveguide (CPW) has become a common experimental tool for studying magnetic films and nanostructures. \cite{Counil, Neudecker, Bilzer, Harward} A ferromagnetic film is placed close to the surface of the CPW  so that a substrate is the furthest outer medium from the CPW. A microwave magnetic field $\tilde{h}$ proportional to the rf current in the central the CPW line excites the precession of the magnetization $\tilde{m}$, which in turn  induces a microwave voltage in CPW. The FMR response is commonly extracted from the reflection ($S_{11}$) or transmission ($S_{21}$) coefficients of scattering parameters using VNA and, hence, the technique is referred to as the VNA-FMR  \cite{Bilzer} or the broadband FMR. \cite{kennewell2010} In practice, only the changes in $S_{21}$ (or $S_{11}$) due to the FMR absorption are of interest and they are detected in a frequency-swept mode \cite{Neudecker}  or in a field-swept mode. \cite{Harward}  It has been proved that a change $\Delta S_{21}$  (or $\Delta S_{11}$) due to microwave absorption of a single ferromagnetic film is proportional to the  complex susceptibility $\chi(\omega)$ or $\chi(H)$. \cite{Counil, Nembach, Ding} The imaginary part of $\Delta S_{21}$ vs. $H$ reflects the Lorentzian curve characteristic of the FMR absorption. The FMR absorption is measured at different frequencies  to determine the effective saturation magnetization $4\pi M_{eff}$, the damping constant $\alpha$ as well as the inhomogeneous contribution to the linewidth $\Delta H_{0}$. \cite{heinrichbland}
In the present paper we rather focus on a less recognized potential of the FMR technique: evaluation of the intensity of the FMR absorption  defined as the integrated FMR  absorption,  which if properly employed, can be used to determine the total magnetic moment. \cite{Celinski}

Interpretation of the VNA-FMR experimental results has been well established. \cite{Bilzer} However, for metallic multilayers or magnetic films \cite{kennewell2007,kennewell2010} in contact with conducting nonmagnetic layers \cite{bailleul} analysis of the experimental data is more complicated.  In opposite to the standard FMR experiments based on microwave cavities with a homogeneous microwave field, in the CPW the microwave field is asymmetric relative a magnetic thin film  and  inhomogeneous due to the shielding of microwaves by the eddy currents.\cite{kennewell2007,kennewell2010} In particular, image currents generated in a floating ground conductor increased a pulsed inductive microwave magnetometer sensitivity, as well as the field strength, resulting in a fourfold increase in overall signal-to-noise ratio. \cite{nibarger}  Recently, Bailleul \cite{bailleul} has shown with the aid of finite-element electromagnetic simulations that the propagation of microwave fields along the CPW is strongly modified when a nonmagnetic film is brought close to it. This effect has been attributed to the shielding of the electric $\tilde{e}$ and/or magnetic field $\tilde{h}$ of the CPW depending on the thickness  of the metallic film.
The shielding is expected to have important consequences for the CPW based VNA-FMR experiments.\cite{bailleul}
For example, it has been reported that the CPW efficiently excites higher order standing spin wave modes across the film with thickness of 30 - 90 nm and  the amplitude of the modes  depends on ordering of FM layers with respect to the CPW.\cite{crew, kennewell2010, kostylev2009, kennewell2010, maksymkostylev}

This paper aims at broadening the above experiments  to ultrathin ferromagnetic layers for which macrospin model is regarded to be fulfilled.  \cite{heinrichbland} For such thin layers (a few nm in thickness) $\Delta S_{21}$ can sometimes be hardly distinguished from the noise. Therefore, any enhancement of the FMR response ($\Delta S_{21}$) is of importance for the VNA-FMR measurements. The purpose of the paper is to investigate the effect of a nonmagnetic buffer layer on the FMR response of  systems that include a stack of ultrathin (buried) exchange decoupled ferromagnetic layers with distinct effective magnetization (magnetic anisotropy) so that the FMR responses of each layer are well separated in the field scale. In particular, we will examine how  the FMR absorption intensity of each magnetic layer depends on the thickness of the conducting buffer layer and on their arrangement with respect to the buffer.

\section{Experimental details}\label{s2}
The multilayer thin films investigated in the present paper by using the VNA-FMR  are intended for spin-transfer oscillators  \cite{Houssameddine} that comprise a [Au/Co]$\times 4$ perpendicular polarizer, an in-plane magnetized [Py/Co] free layer with Permalloy (Py), and an in-plane Co analyzer in contact with a IrMn antiferromagnetic layer.  The composition of the multilayers with the thickness of individual layers in nanometers  is shown in Tab.~\ref{table1}.  The  multilayer films were deposited in a Prevac sputtering system onto the high resistivity ($\rho>2$ k$\Omega$ cm) Si/SiO$_2$ substrates that include [Ti/Au] buffer layers of various thickness.  Since the Ti 4 nm films are used only for an improvement in adhesive strength of our multilayer structures, the [Ti/Au] layers will be referred to as the Au buffers. The base pressure was less than $1\times 10^{-6}$ Pa and the Ar pressure was approximately $10^{-2}$ Pa. All structures were covered with a 5 nm Au cap layer. For the VNA-FMR investigations, the films on the substrates of $19\times 15$ mm were cut to approximately $10\times 15$ mm samples.  The total thickness  of the structures investigated (30 - 100 nm), including the conducting buffer layers, is well below the skin depth at the microwave frequencies of 20 - 30 GHz. Two reference
samples, which comprise a  2.5 nm Co on the Au (10 - 40 nm) and Au (30 - 60 nm) wedge buffers, were deposited in the same conditions.
\begin{table*}
\caption{\label{table1} Composition of multilayer structures which comprise the polarizer (P),  the analyzer (A), and  the free layer (F) separated in between by the Cu spacers. All samples are covered with the Au 5 nm cap layers. The polarizer consists of the Au/Co bilayers repeated four times. The reference samples comprise only the buffer layer, the free Co layer, and the cap layer.}
\begin{tabular}{cccccc}

  \hline
  sample& buffer& \multicolumn{3}{c}{sequence of the  layers  in the stack}&cap layer \\
  \hline \hline
               &               &       P             &    \hspace{1.0cm}    F           &  \hspace{1.1cm}    A    &         \\
  SA\footnotemark[1] &  {Ti\,4/Au\,40} & $\overbrace{(\textrm{Au}\,1/\textbf{Co}\,0.7)_4}$& \hspace{0.3cm}Cu\,4\hspace{0.3cm}$\overbrace{\textbf{Py}\,3/\textbf{Co}\,0.5}$&\hspace{0.3cm}Cu\,3\hspace{0.3cm}$\overbrace{\textbf{Co}\,3/\textrm{IrMn}\,15}$ &\,\,Au\,5\\
  \hline
  &               &       P             &  \hspace{1.0cm}  F           &   \hspace{1.1cm}    A        &     \\
  SA 1\footnotemark[1] &  (Ti\,2/Au\,2)$_5$ \footnotemark[2]&$\overbrace{(\textrm{Au}\,1/\textbf{Co}\,0.7)_4}$& \hspace{0.3cm}Cu\,4\hspace{0.3cm}$\overbrace{\textbf{Py}\,3/\textbf{Co}\,0.5}$&\hspace{0.3cm}Cu\,3\hspace{0.3cm}$\overbrace{\textbf{Co}\,3/\textrm{IrMn}\,15}$ &\,\,Au\,5\\
  \hline

   &               &       P             &          \hspace{1.0cm} F           &   \hspace{1.1cm}    A          &   \\
  SA 2\footnotemark[1] &  (Ti\,2/Au\,2)$_{10}$ & $\overbrace{(\textrm{Au}\,1/\textbf{Co}\,0.7)_4}$& \hspace{0.3cm}Cu\,4\hspace{0.3cm}$\overbrace{\textbf{Py}\,3/\textbf{Co}\,0.5}$&\hspace{0.3cm}Cu\,3\hspace{0.3cm}$\overbrace{\textbf{Co}\,3/\textrm{IrMn}\,15}$ &\,\,Au\,5\\
  \hline
               &               &       A             &      \hspace{1.0cm}    F           &   \hspace{1.1cm}    P        &     \\
  SB & {Ti\,4/Au\,40} & $\overbrace{\textrm{IrMn}\,10/\textbf{Co}\,3}$ &\hspace{0.3cm}Cu\,3\hspace{0.3cm}$\overbrace{\textbf{Co}\,0.5/\textbf{Py}\,3}$ &\hspace{0.4cm}Cu\,4\hspace{0.3cm}$\overbrace{(\textrm{Au}\,1/\textbf{Co}\,0.7)_4}$ &  \,Au\,5\\
   \hline
               &               &                      &  \hspace{1.0cm} F           &   \hspace{1.1cm}    A         &    \\
  SC & {Ti\,4/Au\,10} & - &\hspace{0.3cm}Cu\,3\hspace{0.3cm}$\overbrace{\textbf{Co}\,0.5/\textbf{Py}\,3}$ &\hspace{0.3cm}Cu\,4\hspace{0.3cm}$\overbrace{\textbf{Co}\,3/\textrm{IrMn}\,10}$  &\,Au\,5\\
    \hline            &               & \hspace{0.0cm} A             &    \hspace{1.0cm} F           &   \hspace{1.1cm}   P   &     \\
  SD & {Ti\,4/Au\,10} & $\overbrace{\textrm{IrMn}\,10/\textbf{Co}\,3}$ &\hspace{0.3cm}Cu\,3\hspace{0.3cm}$\overbrace{\textbf{Co}\,0.5/\textbf{Py}\,3}$ &\hspace{0.4cm}Cu\,4\hspace{0.3cm}$\overbrace{(\textrm{Au}\,1/\textbf{Co}\,0.7)_4}$  & \,Au\,5\\
   \hline \hline
          &               &              &    \hspace{1.0cm} F           &  &     \\
  Ref-1& {Ti\,4/Au wedge(10-40)} & - & \hspace{1.0cm}$\overbrace{\textbf{Co}\,\,2.5}$ & \hspace{2.0cm} - \hspace{0.8cm} &Au\,5\\\\
  Ref-2 & {Ti\,4/Au wedge(30-60)} & - & \hspace{1.0cm}\textbf{Co}\,\,2.5 &\hspace{2.0cm} - \hspace{0.8cm} &Au\,5\\
  \hline
\end{tabular}
\footnotetext[1]{Samples SA, SA1 and SA2 have the same structure except buffer layers.}
\footnotetext[2]{The subscripts denote the number of repetition and the other numbers denote thickness in nanometers.}
\end{table*}
\begin{figure}
\includegraphics {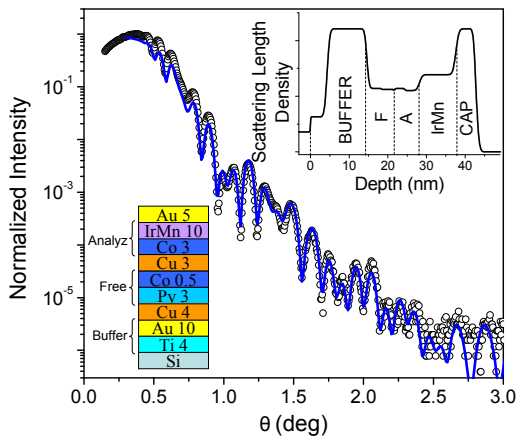}
\caption{\label{fig:1} X-ray reflectivity of the multilayer sample SC that comprises an analyzer and a free layer on a Au buffer layer (lower inset).
For simulation (blue curve) the same parameters were used  as that presented in Tab.~\ref{table1} and the interface roughness of 0.5 - 0.7 nm. The composition profile of the whole structure is shown in the upper inset.}
\end{figure}

Crystalline structure was determined using X-ray diffraction. Diffraction profiles were measured in the Bragg-Brentano geometry and analysis of the diffraction profiles indicates that the Au buffers show a strong (111) texture since only the (111) and (222) peaks were visible. The width of diffraction lines are 0.7 deg, 1 deg, for the free  {Py/Co} layer and IrMn, respectively.  For the 10 and 40 nm thick Au buffer, the width is 1.1 deg and 0.4 deg, respectively. Using the Scherrer formula  we estimate  crystallite size as 9, 25 nm for the 10 and 40 nm Au buffer, respectively. On the basis of a fitting procedure with the use of SimulReflec \cite{simulreflec} software for x-ray reflectivity data the composition profiles of a few chosen multilayers were determined as it is shown in Fig.~\ref{fig:1}, as a typical example. Thickness of the individual layers (Fig.~\ref{fig:1}, the upper inset) is in agreement with that assumed from technological parameters. The roughness estimated from the fitting is of 0.5 - 0.7 nm.
\begin{figure}[h1]
\includegraphics {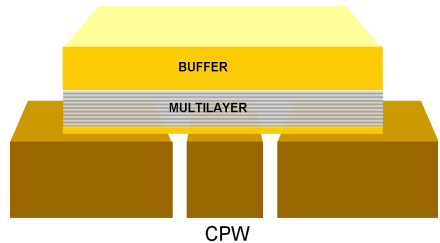}
\caption{\label{fig:2a} A sketch of a piece of coplanar waveguide (CPW) with  typical thin film structure that comprises a magnetic multilayer and a conducting buffer layer. A substrate above the buffer is not shown for clarity. }
\end{figure}

The magnetization reversals of the multilayers were examined using a standard vibrating sample magnetometer at room temperature. The measurements confirmed that the multilayers comprise the perpendicular magnetized polarizer with the effective magnetization $4\pi M_{eff} \approx - 3$ kG, the in-plane magnetized free layer with $4\pi M_{eff} \approx 8$ kG, and the analyzer with $4\pi M_{eff} \approx 12$ kG and the exchange-bias field of 200 Oe. The individual P, F, and A magnetic layers have approximately the same magnetic moments.
\begin{figure}[h1]
\includegraphics {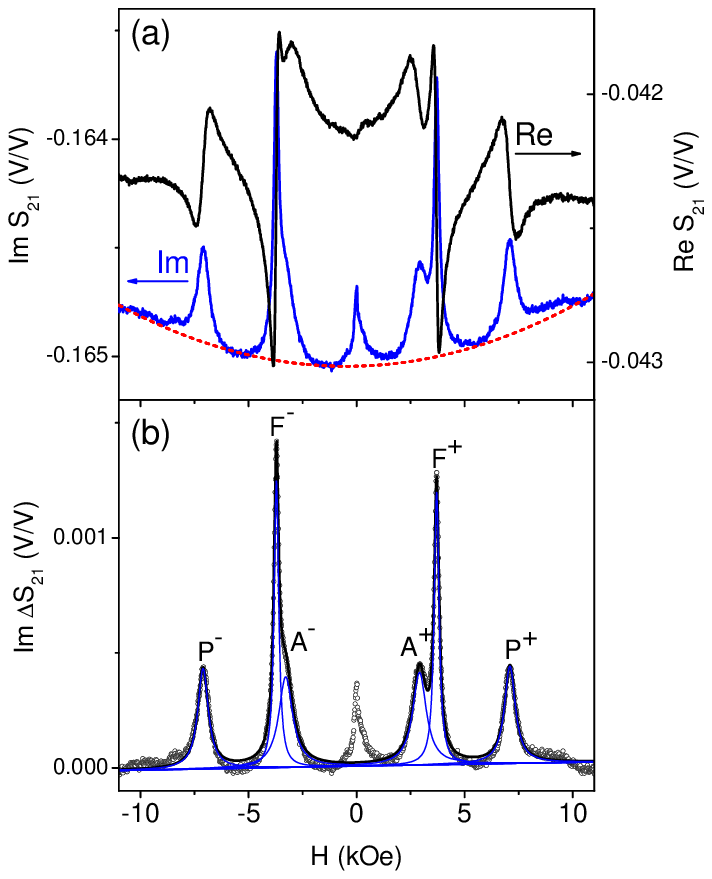}
\caption{\label{fig:2} (a) Typical in-plane VNA-FMR spectrum of the  SA structure with the real Re~$S_{21}$ and  imaginary Im~$S_{21}$ parts of the complex transmission parameter  $S_{21}$. (b) The same spectrum with the values of Im~$\Delta S_{21}$  adjusted by removing a background (red dashed curve). The spectrum was measured with the magnetic field sweep from ~ + 10 kOe to ~ -10 kOe so that the number FMR absorptions is doubled. The central peak at $H=0$ is related to an absorption due to magnetization reversal. Blue lines (in colors - online) show the fits of the spectra with the Lorentzians. }
\end{figure}
\section{CPW VNA-FMR measurement technique}\label{s3}
\begin{figure}
\includegraphics {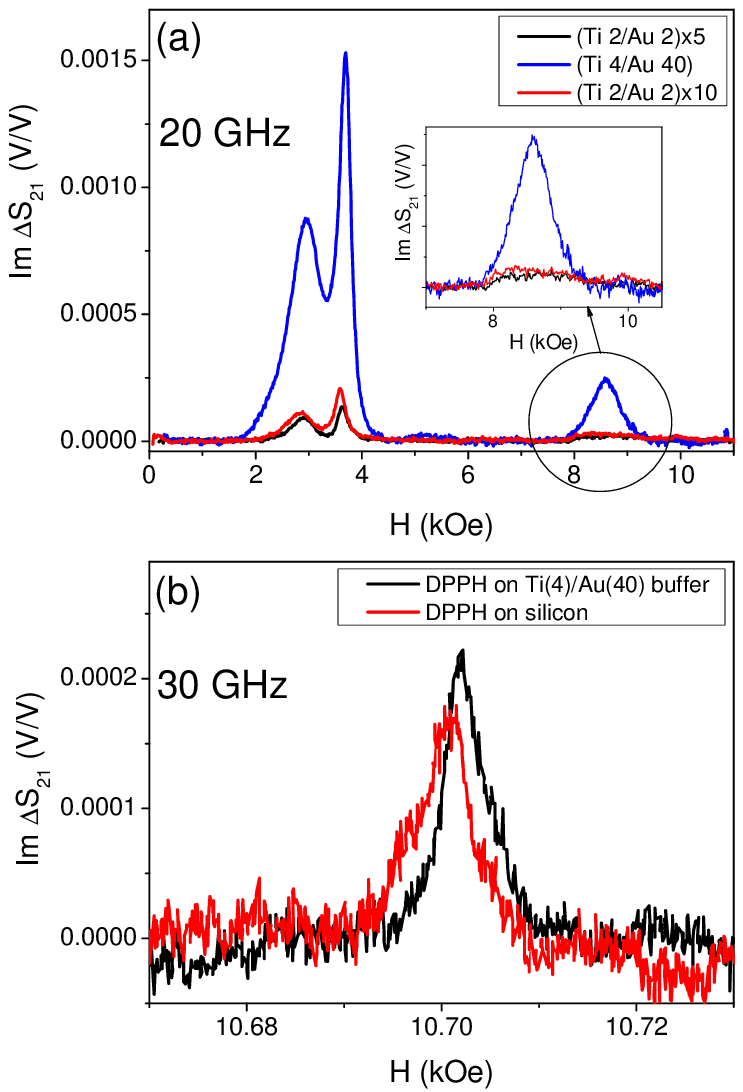}
\caption{\label{fig:3} (a) Comparison of the FMR absorptions in three magnetic structure (SA, SA1, and SA2) in contact with buffer layers of various structures (sheet resistance).  (b) Comparison of EPR signal from a DPPH  "film" on a 40 nm Au buffer and on a bare Si substrate.}
\end{figure}
A broadband FMR spectrometer based on the VNA-FMR technique was used to measure the FMR spectra of multilayers in the in-plane geometry with an external magnetic field applied in the sample plane. An in-plane microwave field with a frequency of 20 - 30 GHz was applied to the sample using a grounded CPW with a 0.45 mm wide central strip. Our CPW design is similar to that of Southwest Microwave Inc. \cite{Southwest} The samples ($10\times 15$ mm) were placed face down on the CPW so that the buffer layer was always the furthest layer in the investigated multilayer structures as it is schematically shown in Fig.~\ref{fig:2a}. The complex transmission parameter $S_{21}$ was measured with the VNA at a fixed frequency - typically 20 GHz - while the external magnetic field was swept between +10, 0, and -10 kOe. Since the FMR signals were measured in two quadrants,  the number of FMR peaks is doubled as it is shown in Fig.~\ref{fig:2}. We find such a measurement procedure helpful in estimating errors in calculations of the area under absorption curve. Figure~\ref{fig:2} (a) shows typical  real part Re $S_{21}$ and  imaginary part Im $S_{21}$ of a complex scattering factor S$_{21} \times exp(i\phi)$ with a phase $\phi$ correction. \cite{kalarickal} Two main features characterize Im $S_{21}$ vs.  $H$: a quasi-parabolic background Im $S_{21}^{0}$ which is related to a nonresonant background of the whole microwave track  and six characteristic absorption peaks (three for positive and three for negative $H$, respectively) plus one central peak. The three peaks P$^{+}$, F$^{+}$, and A$^{+}$ (or, P$^{-}$, F$^{-}$, and A$^{-}$, for the negative field direction) are related to the FMR absorption of the exchange decoupled polarizer, the free layer, and the analyzer, respectively.  The central peak at $H = 0$ Oe is related to a additional absorption due to   magnetization reversal of the F+A structure. The central peak will not be discussed further in this paper.

Figure \ref{fig:2} (a) clearly shows that the experimental data of Im $S_{21}$ can be broken into magnetic and nonmagnetic contributions assuming that a reflection of microwave power is weak in our VNA-FMR set-up. Therefore, following a similar analysis discussed in Refs. \onlinecite{Nembach, Ding}, the complex $S_{21}$ scattering term  may be expressed as
\begin{equation}\label{1}
S_{21}(H,t)\approx S_{21}^{0}(H,t) + \frac{\chi(H)}{\chi_{0}} ,
\end{equation}
where $\chi(H)$ is complex microwave susceptibility, $\chi_{0}$ is complex function of the experimental parameters , such as frequency and film thickness \cite{Ding}. Furthermore, time $t$ takes into account some drift of $S_{21}^{0}$ during measurements.  Assuming that $S_{21}^{0}(H,t)$ depends on $H$ in a nonresonant way and also depends on $t$, we can reasonably approximate $S_{21}^{0}(H,t) \approx A H + B H^{2}$ so that we arrive at a simple relation
\begin{equation}\label{2}
\chi(H) \approx \chi_{0} \Delta S_{21}(H),
\end{equation}
where $\Delta S_{21} = S_{21} - S_{21}^{0}$. Figure \ref{fig:2} (b) shows the measured Im $\Delta S_{21}(H)$) after subtraction of the nonresonant background Im $S_{21}^{0}(H)$. It can be easily shown that the experimental spectrum can be deconvoluted using a set of  Lorentzians. Slight differences in the height and  linewidth $\Delta H$ (FWHM) for P$^{+}$, P$^{-}$ and F$^{+}$,F$^{-}$ peaks serve here as a rough estimate of uncertainties in determination of the $\Delta S_{21}$ absorptions in our VNA-FMR set-up. On the other hand, the substantial difference between A$^{+}$ and A$^{-}$ is due to the unidirectional anisotropy of the analyzer.
Keeping in mind that $\chi^{''}\approx \chi_{0}\; Im \,\Delta S_{21}(H)$, we can further express the area under the FMR absorption peak $I$ as
\begin{equation}\label{3}
 I \propto \int \chi^{''} dH .
\end{equation}
 In theory,  the intensity of the FMR absorption $I$ measured in a microwave cavity  is proportional the total magnetic moment.\cite{gurevich}   However, in this case the FMR intensity studies require a microwave system which can provide reproducible results with a special emphasis on a microwave cavity coupling and a cavity quality factor. \cite{Celinski} In contrast to the discussion in Ref. \onlinecite{Harward},  we have found the magnitude of the FMR absorption ($\propto$ Im~$\Delta S_{21}$) quite stable for the structures of the same size and the same composition. It suggests that we can  compare the  intensities of FMR absorption  of various samples provided that the measurement conditions in the CPW set-up are the same.
\section{VNA-FMR results}\label{s4}
Using our CPW-VNA set-up, we have measured FMR of three SA, SA1, and SA2 structures that comprise identical P, F, and A but  different buffer layers (see Tab.~\ref{table1}, for details). As it is shown in Fig.~\ref{fig:3} (a), the same positions of the resonance field of the P, F, and A layers in the three structures prove that the magnetic layers have the same magnetic properties (e.a., the same values of effective magnetization and the same exchange bias of the analyzer) In contrast,  the signal amplitude of  Im~$\Delta S_{21}$ for SA with the thickest Au buffer is  nearly 6 - 7 times higher than those of  SA1 and SA2. The effect of signal enhancement is even more pronounced for the polarizer P with the perpendicular anisotropy (see the inset in Fig. ~\ref{fig:3} (a)). While the FMR absorptions of the polarizer are barely seen from the noise for the samples SA1 and SA2, the FMR absorption of P in SA is substantial and comparable with those of A and F. As it was shown in our recent paper, \cite{matczak} the enhancement in this case is presumably additionally influenced by a better texture and crystallite size of the polarizer, which was grown on the thick Au 40 nm buffer layer.
Therefore, Fig.~\ref{fig:3} (a) can be regarded as an experimental evidence of shielding of the electromagnetic field in the CPW by a conducting film with a low sheet resistance. It is shown that a highly conducting buffer layer that is the outer conducting layer from CPW can beneficiary affect the excitation of the fundamental mode(s) in our ultrathin film structure. We have checked with a four-point probe that the sheet resistance of SA, SA2, and SA1 buffer layers is of 0.5, 15 and 30 $\Omega$, respectively.

Such a change in the sheet resistance has recently been shown to strongly affect shielding of either $\tilde{h}$ and/or $\tilde{e}$ fields. \cite{bailleul} To check if the field $\tilde{h}$ is really enhanced due to a conducting buffer alone, we performed a similar experiment using DPPH - the common EPR standard compound - dissolved in a nonconducting glue and then deposited on a bare Si substrate and on a Si substrate covered with a Au 40 nm buffer layer, respectively. As it is shown in Fig.~\ref{fig:3} (b), there is no substantial signal enhancement due to the thick Au buffer. Since the DPPH "layers" are insulating and very thick (of 100 $\mu$m), we attribute the FMR signal enhancement for  the  conducting SA, SA1, and SA2 structures to their close proximity of conducting nonmagnetic layers.
\begin{figure}
\includegraphics {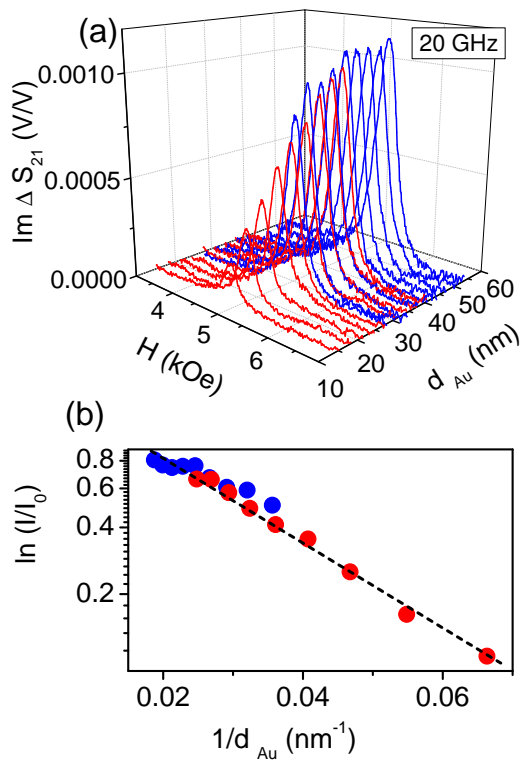}
\caption{\label{fig:4} (a) 3-D plot of FMR absorption of a 2.5nm Co film deposited onto  Ti 4/Au (10 - 40)  (red curves - sample Ref-1) and Ti 4/Au (30 - 60) wedge (blue curves - sample Ref-2), respectively. (b) Log-lin dependence of the normalized  FMR intensity $I/I_{0}$ on the inverse thickness of Au buffer layer.}
\end{figure}

The VNA-FMR measurements of a single ultrathin Co film in contact with a wedged Au buffer  give reference data for quantitative analysis.
For this purpose, we measured Im~$\Delta S_{21}$ for the 2.5 nm Co layer  deposited on the  10 - 40 nm  and Au 30 - 60 nm Au wedges (see Tab.~\ref{table1}). The results of the FMR measurements are shown in Fig.~\ref{fig:4} (a) as a 3-D plot. It is clearly seen that  the amplitude of the FMR absorption of the Co 2.5 nm layer increases  with the thickness of  Au buffer layer and saturates at its thickness of 40-60 nm.  It appears that a dependence of the intensity $I$ vs. $1/d_{Au}$ (see Fig.~\ref{fig:4} (b)) can be approximated by the following expression:
\begin{equation}\label{4}
I =I_{0}\; \exp \;(-\dfrac{d_{o}}{d_{Au}}) = I_{0}\; \exp \; (-\dfrac{R_{sr}}{R^{o}_{sr}}) ,
\end{equation}
where $d_{o}$ is of 38 nm and $I_{0}\approx 1.7$ are the fitting parameters. If we define the sheet resistance of  Au buffer layer  as $R_{sr}=\rho /d_{Au}$ Eq.~(\ref{4}) can be alternatively expressed in terms of the sheet resistance. By assuming the resistivity of bulk gold as 3 $\mu\Omega$cm, $R_{sr}$ of the Au buffer varies from 2 $\Omega$ to 0.33 $\Omega$ for the Au thickness of 10 and 60 nm, respectively. Fitting to the experimental data using Eq.~(\ref{4}) gives $R^{o}_{sr}\approx 0.8$ $\Omega$.

\begin{table}
\caption{\label{table2} Ratio of magnetic moments  and  FMR intensity ratios  of the P, F, and A layers in the SA, SB, and SD structures.}
\begin{tabular}{cccc}

  \hline
  Sample &     $m_{P}:$    &     $m_{F}$ :   &    $m_{A}$    \\
  \hline
	SA,SB,SD&1	&0.8	&1.07	\\
  \hline \hline
	&$I_{P}:$&$I_{F}:$&$I_{A}$\\
	\hline
	SA - inverse structure&1&0.84&0.97\\
	SB - simple structure&1&0.86&0.58\\
	SD - simple structure&1&0.64&0.40\\
	\hline
	
\end{tabular}
\end{table}

\begin{figure*}
\includegraphics {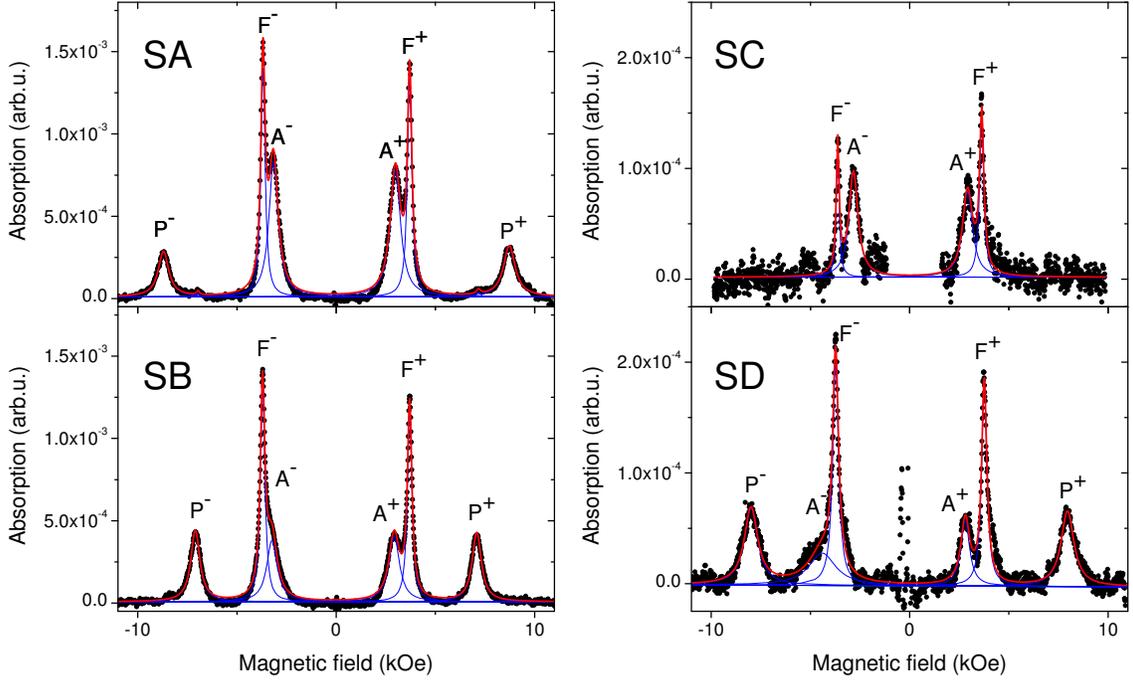}
\caption{\label{fig:5} FMR absorption vs. $H$ of a series of SA - SD multilayer thin film structures prepared for spin-transfer-torque oscillators with a perpendicular polarizer P, a free layer F, and an in-plane analyzer A pinned to IrMn layer. Blue lines (in colors - online) show the fits of the spectra with the Lorentzians. The "central"peak is subtracted for clarity.}
\end{figure*}

 Fig.~\ref{fig:5} shows the  effect of the FMR absorption enhancement  observed in  more complex structures  SA, SB, SC, and SD that include the P, F, and A magnetic layers in various arrangements with respect to the buffer layers (Tab.~\ref{table1}). For some purposes, which are out of scope of the present paper, the sample SC has no polarizer. Comparing Figs. \ref{fig:5} (SA), (SB) with (SC), and (SD), one can see that the FMR amplitudes for the samples SA and SB deposited on  the Au 40 nm buffers are about ten times higher than those of the samples SC and SD deposited onto the Au 10 nm buffers. Besides, a clear decrease in the signal-to-noise ratio is seen in Fig.~\ref{fig:5} for SA and SB in comparison with the SC and SD structures.

As can be seen in Tab.~\ref{table1}, the multilayer samples differ in  sequences of the magnetic layers. The P-F-A and A-F-P structures are referred to as the simple and the inverse structures, respectively. Let us examine the impact of the arrangements of  magnetic layers  on their FMR intensity ratios.   According to Eq.~(\ref{3}), the FMR intensity ratio of $I_{P}$ to $I_{F}$ to $I_{A}$ should be the same as the magnetic moments ratio $m_{P}:m_{F}:m_{A}$  (Tab.~\ref{table2}) provided the dynamic field $\tilde{h}$ is homogeneous.   However,  Tab.~\ref{table2} shows that  $I_{P}:I_{F}:I_{A}$ ratios depend on an arrangement of  magnetic layers. For the simple structures SB and SD the intensity ratios are distinctly different from that of $m_{P}:m_{F}:m_{A}$. Moreover, the intensity of FMR absorption is strongly diminished for the analyzer A with a IrMn layer in close contact with the Au buffer. On the other hand, for the inverse structure SA the FMR intensity ratio does not differ much from the ratio of the magnetic moments.

\section{Discussion}\label{s5}
\begin{figure}
\includegraphics {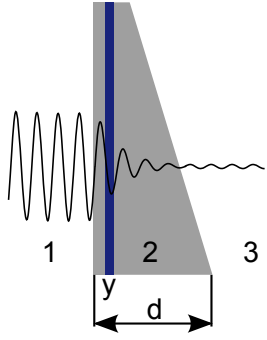}
\caption{\label{fig:6} Sketch showing a structure that comprise an ultrathin Co  film and a Au wedge $10<d<50$ nm. Microwaves are partially reflected and transmitted through the wedge.}
\end{figure}

Discussion of our experimental data is based on the essential results of Ref.~\onlinecite{kostylev2009}. (\textit{i}) In contrast to the common cavity FMR measurements, a conducting thin film sample in the CPW - FMR is illuminated by microwaves asymmetrically from the front surface of the film as it is shown in Fig.~\ref{fig:6}. (\textit{ii}) In such a geometry, the thin film sample with a thickness $d$ less than the skin depth  the microwave magnetic field decays more strongly than exponentially. (\textit{iii}) In a highly conducting film the microwave magnetic field is strongly inhomogeneous.

We can write the scattering parameter $S_{21}$  in terms of the complex reflection coefficient $\Gamma$  and the complex propagation factor $\gamma_{0} \gamma_{f}$.\cite{Bilzer}
\begin{equation}\label{5}
\frac{S_{21}}{S_{21}^{0}}=\frac{{\Gamma}^2 - 1}{{\Gamma}^2  \exp({-\gamma_{0} \gamma_{f} \delta})-\exp({\gamma_{0} \gamma_{f} \delta})}.
\end{equation}
$\gamma_{0}=i\omega / \upsilon_{ph}$ is the complex propagation factor of the unloaded CPW, where $\omega$ is the angular frequency of microwave field and  $\upsilon_{ph}$  is the phase velocity of microwaves in the the CPW.  $\gamma_{f}$ is the propagation index of the loaded CPW.
$\delta$ is the film width and $S_{21}^{0}$ is the scattering parameter  of the empty CPW. Keeping only linear term in the expansion of Eq.~(\ref{5}) and assuming that $|\Gamma|\ll 1$,\cite{kennewell2010} we obtain
\begin{equation}\label{6}
\frac{S_{21}}{S_{21}^{0}}=\exp({-\gamma_{0} \gamma_{f} \delta}).
\end{equation}
The propagation index $\gamma_{f}$ can be further approximated  in terms of the characteristic impedance $\gamma_{f}=\sqrt{\frac{Z_{0}-Z_{r}}{Z_{0}}}\approx 1+\frac{Z_{r}}{2Z_{0}}$, (Eq.~(13) in Ref.~\onlinecite{kostylev2009}), where $Z_{0}$ is the characteristic impedance of the unloaded CPW and $Z_{r}$ is the surface impedance of a thin film placed on the CPW. Hence,
\begin{equation}\label{7}
\frac{S_{21}}{S_{21}^{0}}=\exp \left({-\gamma_{0} \frac{Z_{r}}{2Z_{0}} \delta}\right).
\end{equation}
According to  Ref.~\onlinecite{bailleul}
\begin{equation}\label{8}
Z_{r}=R_{sr} \frac{\delta}{w}=\frac{\rho}{d} \frac{\delta}{w},
\end{equation}
where $R_{sr}$ is the sheet resistance and $w$ is a width of the central line of CPW (see. Fig.~2 in Ref.~\onlinecite{bailleul}). Eventually, in agreement with Ref.~\onlinecite{kostylev2009}, we can express the measured scattering coefficient $S_{21}$ in terms of geometrical parameters of CPW and the film placed on it
\begin{equation}\label{9}
\frac{S_{21}}{S_{21}^{0}}\propto \exp \left( {- \frac{\frac{\gamma_{0} \rho \delta^2 }{2 Z_{0} w}}{d}}\right)\propto \exp \left({-\frac{d_{0}}{d}}\right),
\end{equation}
which has the same form as the fitting formula Eq.~(\ref{4}) to the experimental data shown in Fig.~\ref{fig:4}.
Let us estimate $d_{0}$. For $Z_{0}=25-50$ $\Omega$, $|\gamma_{0}|=\omega / \upsilon_{ph}=7$ cm$^{-1}$, $\delta=0.4$ cm,
$\upsilon_{ph}=1.8\times10^{10}$ cm/s, and $\omega /2\pi=20$ GHz  the estimated range of $d_{0}$ is between 8 and 15 nm if we assume the resistivity $\rho=3$ $\mu \Omega$cm of the gold buffer the same as for bulk. In practice, the resistivity of several nanometers thick gold films is several times higher \cite{Sambles} so that $d_{o}=38$ nm estimated from fitting of the experimental data (Fig.~\ref{fig:4}) according to Eq.~(\ref{4}) is in agreement with the above model.

Let us consider an ultrathin Co 2.5 nm film  deposited on a gold wedged buffer layer of thickness $10<d<50$ nm as it shown in Fig.~\ref{fig:6}. The Co film  plays the role of a tag useful for monitoring of the dynamic magnetic field $\tilde{h}$. Since the Co film is very thin in comparison to the Au buffer, we assume that the entire structure has the conductivity of gold wedge (region 2). A transverse wave with a wavenumber $k_{1}$  and with the amplitude equal to unity incidents perpendicular to the film surface from region 1 with the permittivity $\epsilon _{1}$ the same as for region 3. \cite{kostylev2009} The permittivity of region 2 $\epsilon _{2}$ is complex. Taking continuity boundary conditions at the boundaries of regions 1, 2, and 3 for $\tilde{h}_{x}$ and $\tilde{e}_{z}$ we have

\begin{eqnarray}
  \tilde{h}_{x\,1} &=& \exp(-i k_{1} y)+B_{1}\exp(i k_{1} y), \\
  \tilde{h}_{x\,2} &=& A_{2}\exp(-i k_{1}y)+B_{2}\exp(i k_{1} y), \\
  \tilde{h}_{x\,3} &=& A_{3}\exp(-i k_{1}y)
\end{eqnarray}

with
\begin{eqnarray}
  B_{1} &=& \frac{(r-1) (\exp(2ik_{2}d)-1)}{D}, \\
  A_{2} &=& \frac{2(1+\sqrt r)\exp(2ik_{2}d)}{D}, \\
  B_{2} &=& \frac{2(\sqrt r-1)}{D}, \\
  A_{3} &=& \frac{4\sqrt r\exp(i k_{2}d(1+\sqrt r))}{D},
\end{eqnarray}
where $r=\epsilon_{1}/\epsilon_{2}$ and $D=\exp(2ik_{2}d)(1+\sqrt r)^2-(1-\sqrt r)^2$. Since $\epsilon_{1}=1$ and $\epsilon_{2}$ of gold is imaginary and very large,\cite{kennewell2010} $\mid r\mid \ll 1$ so that after expanding exponential functions and keeping only the linear terms of expansions we obtain that $\tilde{h}$ varies linearly with $y$ as
\begin{equation}\label{11}
\tilde{h}_{x\,2}\approx 2 \frac{d-y}{d},
\end{equation}
where 2 on the right-hand side denotes that the amplitude at the front of the structure is doubled due to the positive interference.
\begin{figure}[b]
\includegraphics  {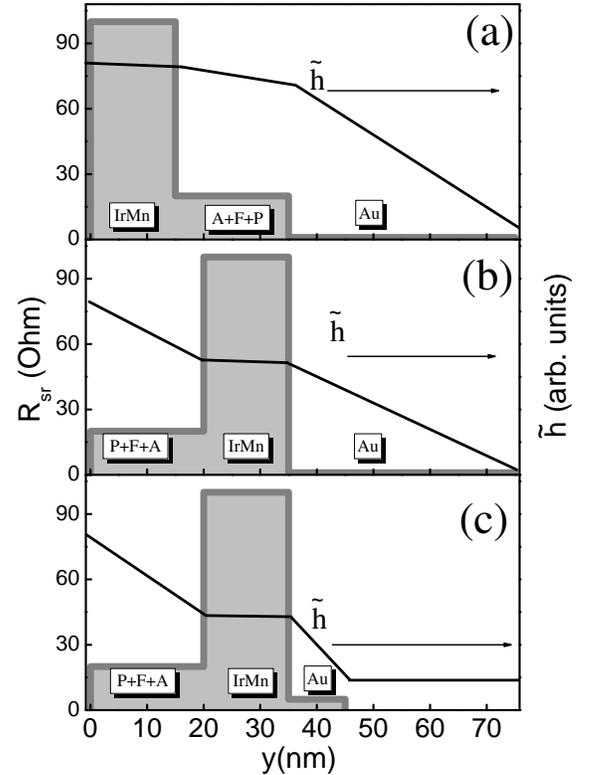}
\caption{\label{fig:7} Distribution of the dynamic magnetic field $\tilde{h}$ inside SA (a), SB (b), and SD (b) structures with diverse sheet resistance $R_{rs}$ of individual layers.}
\end{figure}

In order to extend the model of inhomogeneous dynamic magnetic field within  more complicated structure that comprise P+F+A layers and the Au buffer layer, let us compare the FMR intensity of the CPW-FMR responses shown in Fig.~\ref{fig:5}. In contrast to the single ultrathin Co film on the Au buffer layer, the P+F+A ferromagnetic structure is more extended (of $\sim 20$ nm) and consists of the exchange decoupled (the Cu spacers are 3 - 4 nm thick) Co and Permalloy layers with diverse effective anisotropies. This makes possible observation of the well separated the FMR absorptions of each layer. Exact calculations of the electromagnetic field distribution in such structures would require a set of many boundary conditions \cite{maksymkostylev} with several material parameters. However, we make use  of Eq.~(\ref{11}) taking into account that $\tilde{h}$ in the multilayer is a linear combination of $\tilde{h}$ in individual layers and its slope scales with a sheet resistance ($\rho/d$) of the individual layers. \cite{kennewell2010} Hence, the lower $R_{rs}$ the higher is the slope of $\tilde{h}$ within a layer. $R_{rs}$ values of the individual layers in the entire stack are quite diverse - $R_{rs}$ is the highest for the IrMn layer \cite{umetsu} and the lowest for the Au buffers. Possible distributions of the dynamic field $\tilde{h}$ (black lines) are shown in Fig.~\ref{fig:7} (a), (b), and (c) for  SA, SB, and SD structures, respectively. In other words, we assume in accordance with Eq.~(\ref{11}) that $\tilde{h}(0)=2$ and $\tilde{h}(d)=0$ what is generally not true for very thin stacks (see Fig.~6 in Ref.~\onlinecite{kostylev2009}) so that the microwaves can be partially transmitted out of the stack as it is shown in Fig.~\ref{fig:7} (c). Nonetheless, the distributions of $\tilde{h}$ just depict graphically that its slope in the P+F+A stacks is the highest for the SD sample and the lowest for the SA sample so that the magnetic field inside the SA sample is the most homogeneous. It is also seen that the position of antiferromagnetic IrMn pinning layer  plays an important role in the distribution of $\tilde{h}$ because its resistivity \cite{umetsu} is about 100 times higher than that of Au. \cite{Sambles}  Hence, in accordance with the sketches shown in Fig.~\ref{fig:7}, the dynamic magnetic field is the most inhomogeneous  for the SD sample with the simple P+F+A structure. In contrast, for  the SA  sample with the inverse A+F+P structure the FMR intensity ratio  is in a fairly good agreement with the ratio of  magnetic moments estimated from geometry of P, F, and A layers.

\section{Conclusions}\label{s6}
We have expanded former \cite{Counil,Neudecker,Bilzer,Harward, kennewell2010, Nembach, Ding, kennewell2007, crew, kostylev2009} coplanar waveguide based VNA-FMR studies of thin magnetic films  to ultrathin magnetic structures deposited on the buffer layers with diverse sheet resistance $R_{sr}$. We showed that the intensity of the  FMR absorption of the single ultrathin Co layer  depends on the thickness $d_{Au}$ of the conducting Au buffer $\propto \exp(-d_{0}/d_{Au})$ or, equivalently, on the buffer sheet resistance  $\propto \exp(-R_{sr}/R^{0}_{sr})$. We showed that the measured FMR absorption intensities of  structures composed of several exchange decoupled ultrathin magnetic layers do not scale in proportion to their magnetic moments as would be expected. On the contrary, the ratios  of  FMR absorption intensity of the individual P, F, and A layers depend on their arrangement with respect of the buffer layer. The above mentioned findings are interpreted in terms of the microwave shielding effect by the conducting nonmagnetic buffers and the inhomogeneous dynamic field $\tilde{h}$. The coplanar waveguides (micro-antennas) are widely used in numbers of spintronic devices \cite{grundler} and the  enhancement of FMR response has potential to be applied in spintronic devices.

\section*{Acknowledgment} This research has been conducted in framework of Project NANOSPIN PSPB-045/2010 supported by a grant from
Switzerland through  Swiss contribution to the enlarged European Union.  The authors thanks  Dr B. Szyma\'nski  for assistance with x-ray measurements, A. Krysztofik for  assistance with FMR measurements, and Dr P. Bala\^z for his help with some calculations.

\end{document}